\documentclass[aip,twocolumn]{revtex4}

\usepackage{amsmath}
\usepackage{amssymb}
\usepackage{stmaryrd}

\usepackage{graphicx}

\usepackage{epsfig}

\def\ov#1{\overline{#1}}

\def\wt#1{\widetilde{#1}}

\def\pd#1#2{\frac{\partial #1}{\partial #2}}

\def\wh#1{\widehat{#1}}

\def\cal#1{\mathcal{#1}}

\newcommand{\bc}{\begin{center}}
\newcommand{\ec}{\end{center}}
\newcommand{\bt}{\begin{tabbing}}
\newcommand{\et}{\end{tabbing}}
\newcommand{\be}{\begin{eqnarray*}}
\newcommand{\ee}{\end{eqnarray*}}
\newcommand{\bs}{\begin{slide}}
\newcommand{\es}{\end{slide}}

\begin{document}

\title{Action-angle coordinates for motion in a straight magnetic field with constant gradient}

\author{Alain J.~Brizard}
\affiliation{Department of Physics, Saint Michael's College, Colchester, VT 05439, USA}

\begin{abstract}
The motion of a charged particle in a straight magnetic field ${\bf B} = B(y)\,\wh{\sf z}$ with a constant perpendicular gradient is solved exactly in terms of elliptic functions and integrals. The motion can be decomposed in terms of a periodic motion along the $y$-axis and a drift motion along the $x$-axis. The periodic motion can be described as a particle trapped in a symmetric quartic potential in $y$. The canonical transformation from the canonical coordinates $(y,P_{y})$ to the action-angle coordinates 
$(J,\theta)$ is solved explicitly in terms of a generating function $S(\theta,J)$ that is expressed in terms of Jacobi elliptic functions. The presence of a weak constant electric field ${\bf E} = E_{0}\,\wh{\sf y}$ introduces an asymmetric component to the quartic potential, and the associated periodic motion is solved perturbatively up to second order.
\end{abstract}

\date{\today}

\maketitle

\section{Introduction}

The motion of a charged particle in a straight magnetic field ${\bf B} = B(y)\,\wh{\sf z}$ with constant perpendicular gradient was considered recently \cite{Brizard_2017} in order to explore the validity of the guiding-center approximation \cite{Cary_Brizard_2009} in the present of strong gradients. There, a single orbit in the $(x,y)$ plane was solved exactly in terms of elliptic functions and integrals \cite{footnote}, and the orbit-averaged position $\langle y\rangle$ and drift velocity $\langle\dot{x}\rangle$ were compared, respectively, with the polarization shift and magnetic-drift velocity predicted by guiding-center Hamiltonian theory \cite{Cary_Brizard_2009}. An excellent agreement was found between exact analytical results and guiding-center predictions, even in the presence of strong magnetic gradients.

The purpose of the present paper is to consider all orbit types associated with the straight magnetic field ${\bf B} = B_{0}\,(1 - y/L)\,\wh{\sf z}$ considered in Ref.~\cite{Brizard_2017}, where $B_{0}$ is the field magnitude at $y = 0$ and $|\nabla\ln B(y)|_{y = 0} \equiv 1/L$ defines a constant gradient length scale $L$. The equations of motion in the $(x,y)$-plane perpendicular  to the magnetic field are 
\begin{equation}
\left. \begin{array}{rc l}
\ddot{x} &=& \Omega_{0}\,(1 - y/L)\,\dot{y} \\
\ddot{y} &=& -\,\Omega_{0}\,(1 - y/L)\,\dot{x}
\end{array} \right\},
\label{eq:xy_exact}
\end{equation}
where $\Omega_{0} = q B_{0}/M$ denotes the constant gyrofrequency for a particle of charge $q$ and mass $M$. By using the canonical momentum conservation law 
\begin{equation}
P_{x} = M\,\left[ \dot{x} \;-\frac{}{} \Omega_{0} \left(y - y^{2}/2L\right)\right] \equiv  M\Omega_{0}L\,\left(u - \frac{1}{2}\right), 
\label{eq:canonical_momentum}
\end{equation}
we can now express the $y$-dynamics in terms of the energy conservation law $E = M\dot{y}^{2}/2 + V(y,u)$, where the potential energy $V(y,u) \equiv M\dot{x}^{2}/2$ represents a quartic potential in $y$, with minima $(\dot{x} = 0)$ at 
$y/L = 1 \pm \sqrt{2u}$ and a local maximum at $y = L$, where $V(L,u) = \frac{1}{2}\,M\,(u\,\Omega_{0}L)^{2}$.

The study of charged particle motion in the presence of a magnetic neutral sheet in a magnetized space plasma has been carried out elsewhere \cite{Speiser_1965,Rothwell_Yates_1984,Parks_2003}. In the present paper, we expand the recent work of Kabin \cite{Kabin_2021} on explicit solutions of this system and calculate, for each orbit type, the action-angle coordinates and the generating function for the canonical transformation to these coordinates. Because the associated 2D orbits are labeled by two exact invariants (energy and canonical momentum along the $x$-axis), the motion is exactly integrable and the construction of the magnetic moment is not required.

The remainder of the paper is organized as follows. In Sec.~\ref{sec:orbits}, the equations of motion \eqref{eq:xy_exact} in the $(x,y)$ plane are transformed into the problem of a particle moving in a symmetric quartic potential, whose solutions in terms of the Jacobi elliptic functions and integrals are well known \cite{Reichl_1984,Brizard_Westland_2017}.   In Sec.~\ref{sec:action}, these orbital solutions are used to calculate an action integral $J \equiv (1/2\pi) \oint P_{y}\,dy$ for each particle orbit in terms of elliptic integrals. In addition, orbital periods are recovered from these action integrals in accordance to standard definitions \cite{Goldstein_2002}. In Sec.~\ref{sec:canonical}, an angle coordinate $\theta$ is introduced for each orbit and these angles are shown to be canonical to the action coordinates introduced in Sec.~\ref{sec:action}. In Sec.~\ref{sec:generating}, the function $S$ that generates the canonical transformation $(P_{y},y) \rightarrow (J,\theta)$ is derived from the relation \cite{Arnold_1989} $P_{y}\,dy = J\,d\theta + dS$. In Sec.~\ref{sec:electric}, we investigate the problem of a particle moving in an asymmetric quartic potential in the presence of a constant electric field. Finally, our work is summarized in Sec.~\ref{sec:summary}, while useful definitions of the Jacobi epsilon and zeta functions are presented in App.~A.

\section{\label{sec:orbits}Particle orbits in a straight magnetic field with constant gradient}

\begin{figure}
\epsfysize=1.6in
\epsfbox{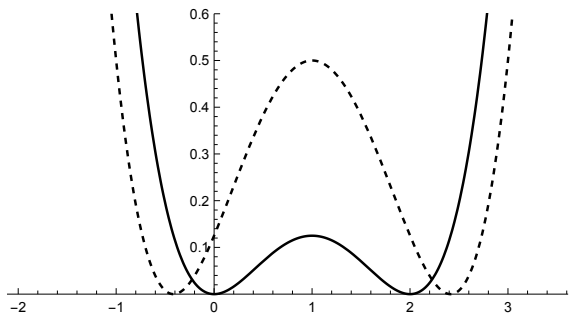}
\caption{Plot of the quartic potential \eqref{eq:V_y} in the range $-2 \leq y \leq 3$ for $u = 1/2$ (solid) and $u = 1$ (dashed). The two wells (with minima at at $y = 1 \pm \sqrt{2u}$) are separated by a barrier (local maximum) at $y = 1$ with height $u^{2}/2$.}
\label{fig:V_y}
\end{figure}

In this Section, we derive solutions to all possible orbits associated with the equations of motion \eqref{eq:xy_exact}. We begin with their dimensionless form:
\begin{eqnarray}
x^{\prime\prime} &=& (1 - y)\,y^{\prime}, \label{eq:x_pp} \\
y^{\prime\prime} &=& -\,(1 - y)\,x^{\prime}, \label{eq:y_pp} 
\end{eqnarray}
where the coordinates $(x,y)$ are normalized to the length scale $L$ and time derivatives (denoted by primes) have been normalized to the dimensionless time $\tau = \Omega_{0}t$. Equations \eqref{eq:x_pp}-\eqref{eq:y_pp}  have two conservation laws: the total (kinetic) energy
\begin{equation}
\left(x^{\prime}\right)^{2} \;+\; \left(y^{\prime}\right)^{2} \;=\; v^{2},
\label{eq:kinetic}
\end{equation}
and the canonical momentum along the $x$-axis
\begin{equation}
x^{\prime} \;=\; u \;-\; \frac{1}{2}\,\left(1 - y\right)^{2},
\label{eq:momentum}
\end{equation}
where the normalized perpendicular kinetic energy $v^{2}$ and the normalized shifted canonical $x$-momentum $u > 0$ are constants. Kabin \cite{Kabin_2021} considered the action integrals for periodic motion along the $y$-axis representing a particle moving in the quartic potential 
\begin{equation}
V(y,u) \;\equiv\; \frac{1}{2}\,(x^{\prime})^{2} \;=\; \frac{1}{8} \left[2 u \;-\frac{}{} (1 - y)^{2}\right]^{2}, 
\label{eq:V_y}
\end{equation}
with total energy $E = v^{2}/2$. This quartic potential, as shown in Fig.~\ref{fig:V_y}, is composed of two wells with minima at $y = 1 \pm \sqrt{2u}$, which are separated by a barrier (local maximum) at $y = 1$ with height $u^{2}/2$. By inserting the momentum conservation law \eqref{eq:momentum} into the equation of motion \eqref{eq:y_pp}, we obtain the second-order ordinary differential equation
\[ y^{\prime\prime} \;=\; -\,u\,(1 - y) \;+\; \frac{1}{2} (1 - y)^{3}, \]
which can be transformed into the standard Duffing equation, whose solution in terms of the Jacobi elliptic functions is well known \cite{Nayfeh_1973,Lawden_1989}. In the remainder of this section, we will use these elliptic solutions to solve the problem of finding all possible orbits associated with the coupled equations of motion \eqref{eq:xy_exact}.

First, using the transformation  $s \equiv 1 - y$ and the definition $v \equiv u\,{\sf e}$, Eq.~\eqref{eq:y_pp} is transformed into the energy conservation law
\begin{equation}
\left(s^{\prime}\right)^{2} \;=\; \left( u\,(1 + {\sf e}) \;-\; \frac{1}{2}\,s^{2} \right) \left( u\,({\sf e} - 1) \;+\; \frac{1}{2}\,s^{2}\right).
\label{eq:s_prime}
\end{equation}
The solution for $s(\tau)$ is periodic, with two turning point $\pm s_{1}$, where $s_{1} = \sqrt{2u\,(1 + {\sf e})}$ for all ${\sf e}$. If ${\sf e} < 1$, however, two additional turning points appear at $\pm s_{2}$, where $s_{2} = \sqrt{2u\,(1 - {\sf e})}$. In what follows, the initial conditions used to integrate Eqs.~\eqref{eq:x_pp}-\eqref{eq:y_pp} are $(x_{0},x_{0}^{\prime}) = (0, -\,u\,{\sf e})$ and $(y_{0},y_{0}^{\prime}) = (1 - s_{1}, 0)$, which are consistent with the conservation laws \eqref{eq:kinetic}-\eqref{eq:momentum}.

\subsection{Type I Orbit}

\begin{figure}
\epsfysize=2.8in
\epsfbox{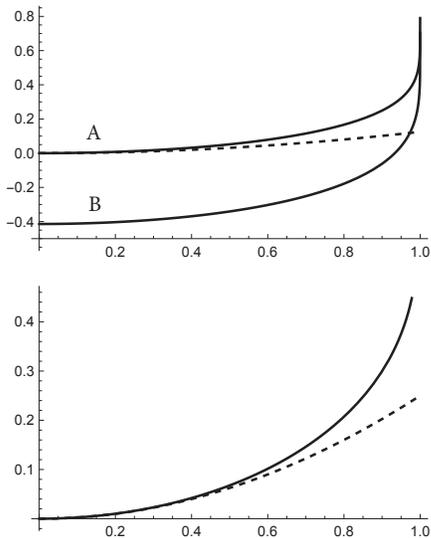}
\caption{(Top) Plot of the orbit-averaged position $\langle y\rangle(u,{\sf e})$ in the range $0 \leq {\sf e} \leq 1$ for $u = 0.5$ (A) and 1.0 (B). At the separatrix $({\sf e} = 1)$, the orbit-averaged position is $\langle y\rangle(u,1) = 1$ for  all $u > 0$. (Bottom) Plot of the normalized orbit-averaged drift velocity $\langle x^{\prime}\rangle/u$ in the range $0 \leq {\sf e} \leq 1$. In both plots, the dashed curves represent the guiding-center approximations discussed in Ref.~\cite{Brizard_2017}.}
\label{fig:y_average}
\end{figure}

When ${\sf e} < 1$, the particle is either trapped in the left well $1 - s_{1} < y < 1 - s_{2}$ or the right well $1 + s_{2} < y < 1 + s_{1}$, i.e., $y$ does not go through $1$ (Kabin \cite{Kabin_2021} calls these type I orbits). We now consider periodic solutions for $s(\tau) = s_{1}\,P(\nu\tau)$, expressed in terms of solutions to the Jacobi elliptic differential equation
\begin{equation}
\left(P^{\prime}\right)^{2} \;=\; \left(1 - P^{2}\right) \left( P^{2} \;-\; m^{\prime} \right),
\label{eq:P_prime}
\end{equation}
where $0 < m^{\prime} = (1 - {\sf e})/(1 + {\sf e}) = s_{2}^{2}/s_{1}^{2} < 1$. For this case \cite{Lawden_1989,NIST_Jacobi}, the solution of Eq.~\eqref{eq:P_prime} is $P(\zeta) = {\rm dn}(\zeta|m)$, where $\zeta = \nu\tau$ and $\nu = \sqrt{u\,(1 + {\sf e})/2}$, with $m = 1 - m^{\prime} = 2\,{\sf e}/(1 + {\sf e}) < 1$. Hence, the solution for a type-I orbit is
\begin{equation}
y(\tau; u, {\sf e}) \;=\; 1 \;-\; s_{1}\;{\rm dn}(\zeta\,|\, m),
\label{eq:y_4}
\end{equation}
where the orbit is trapped in the left well (see Fig.~\ref{fig:V_y}): $1 - s_{1} \leq y \leq  1 - s_{2}$. The particle velocity in this case is
 \begin{equation}
 y^{\prime}(\tau; u, {\sf e}) \;=\; 2\,u\,{\sf e}\;{\rm cn}(\zeta\,|\, m)\,{\rm sn}(\zeta\,|\, m),
 \label{eq:y_I_velocity}
 \end{equation}
and the period for this motion (normalized to $\Omega_{0}$) is $T = 2\,{\sf K}(m)/\nu$, where 
\[ {\sf K}(m) \;\equiv\; \int_{0}^{\pi/2}\frac{d\varphi}{\sqrt{1 - m\,\sin^{2}\varphi}} \]
denotes the complete elliptic integral of the first kind \cite{footnote}. We can now calculate the orbit-averaged position $\langle y\rangle \equiv \int_{0}^{T} y(\tau)\,d\tau/T$ using Eq.~\eqref{eq:y_4}, which can be expressed as
\begin{eqnarray}
\langle y\rangle(u,{\sf e}) &=& 1 \;-\; \frac{\nu}{{\sf K}(m)}\int_{0}^{2{\sf K}(m)} {\rm dn}(\zeta|m)\,d\zeta \nonumber \\
 &=& 1 \;-\;  \frac{\pi\nu}{{\sf K}(m)}.
\end{eqnarray}
Figure \ref{fig:y_average} (top) shows the orbit average $\langle y\rangle(u,{\sf e})$ in the range $0 \leq {\sf e} \leq 1$ for $u = 0.5$ (A) and 1.0 (B). For all $u > 0$, the orbit average is $\langle y\rangle(u,1) = 1$. The dashed curve in the top frame of Fig.~\ref{fig:y_average} shows the guiding-center approximation for the polarization shift calculated in Ref.~\cite{Brizard_2017}. We note that the comparison of predictions from guiding-center theory \cite{Cary_Brizard_2009} with the exact solution is possible because the particle orbit never crosses the neutral sheet (where the magnetic field vanishes) at $y = 1$.

\begin{figure}
\epsfysize=3in
\epsfbox{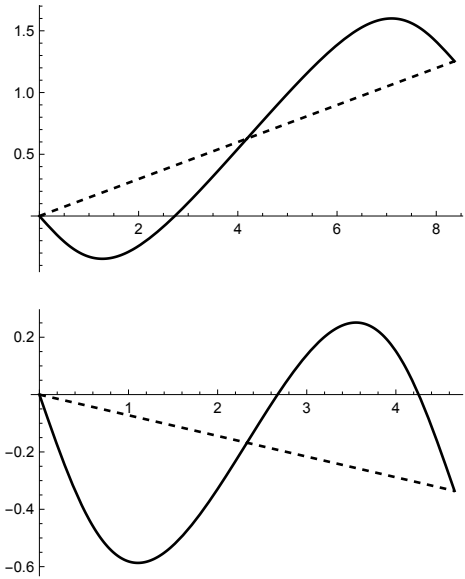}
\caption{(Top) Plot of $x(\tau;u,{\sf e})$ for $u = 1/2$ and ${\sf e} = 0.9$ (solid curve), with the drift motion $\langle x^{\prime}\rangle\,\tau$ shown as a dashed line. (Bottom) Plot of $\ov{x}(\tau;u,{\sf e})$ for $u = 1/2$ and ${\sf e} = 1.8$ (solid curve), with the drift motion $\langle 
\ov{x}^{\prime}\rangle\,\tau$ shown as a dashed line.}
\label{fig:x_drift}
\end{figure}

From the solution \eqref{eq:y_4} for $y(\tau; u, {\sf e})$, we can also obtain an expression for $x^{\prime}(\tau; u, {\sf e})$ from Eq.~\eqref{eq:momentum}:
\begin{equation}
x^{\prime}(\tau;u,{\sf e}) \;=\; u \left[ 1 \;-\frac{}{} (1+{\sf e})\,{\rm dn}^{2}(\nu\tau|m) \right].
\label{eq:x_I}
\end{equation}
First, when averaging the right side of Eq.~\eqref{eq:x_I} over an entire orbital period $T = 2{\sf K}/\nu$, we find the orbit-averaged drift velocity
\begin{eqnarray}
\left\langle x^{\prime}\right\rangle(u,{\sf e}) &=& u \left[ 1 \;-\; \frac{1+{\sf e}}{2\,{\sf K}}\int_{0}^{2{\sf K}} {\rm dn}^{2}(\zeta|m)\,d\zeta \right] \nonumber \\
 &=& u \left[ 1 \;-\; (1 + {\sf e})\;\frac{{\sf E}(m)}{{\sf K}(m)} \right],
\end{eqnarray}
where ${\sf E}(m) = \int_{0}^{\pi/2}\sqrt{1 - m\,\sin^{2}\varphi}\,d\varphi$ denotes the complete elliptic integral of the second kind. The orbit-averaged drift velocity is shown in Fig.~\ref{fig:y_average} (bottom) in addition to its guiding-center approximation \cite{Brizard_2017} shown as a dashed curve. Next, the explicit integration of Eq.~\eqref{eq:x_I}, using the initial condition $x(0) = 0$, yields the expression
\begin{eqnarray}
x(\tau;u,{\sf e}) &=& u\,\tau \;-\; 2\nu \int_{0}^{\nu\tau} {\rm dn}^{2}(\zeta|m)\,d\zeta \nonumber \\
 &=&  u\,\tau \;-\; 2\nu\;{\cal E}(\nu\tau|m) \nonumber \\
 &\equiv& \left\langle x^{\prime}\right\rangle\,\tau \;-\; 2\nu\;{\cal Z}(\nu\tau,m),
\end{eqnarray}
where the $2{\sf K}$-periodic Jacobi zeta function ${\cal Z}(\nu\tau,m)$ is expressed in terms of the quasi-periodic Jacobi epsilon function ${\cal E}(\nu\tau|m)$ according to the expression \cite{NIST_Jacobi}: ${\cal Z}(\zeta,m) = {\cal E}(\zeta|m) - {\sf E}(m)\zeta/{\sf K}(m)$; see App.~\ref{sec:App} for details about the Jacobi epsilon and zeta functions. A plot of $x(\tau;u,{\sf e})$ is shown in the top frame of Fig.~\ref{fig:x_drift} for $u = 1/2$ and ${\sf e} = 0.9$ in the range $0 \leq \tau \leq 2{\sf K}/\nu$, with the drift motion $\langle x^{\prime}\rangle\,\tau$ shown as a dashed line.

\subsection{Type II Orbit}

When ${\sf e} > 1$, the particle is trapped in the well $1-s_{1} < y < 1+s_{1}$ above the potential barrier (see Fig.~\ref{fig:V_y}) at $y = 1$ (Kabin \cite{Kabin_2021} calls these type II orbits). Since $1 < m = 2{\sf e}/(1 + {\sf e}) < 2$, the orbit solution can be expressed using the identity  $(m > 1)$: ${\rm dn}(\zeta|m) = {\rm cn}(\sqrt{m}\,\zeta|m^{-1}) \equiv {\rm cn}(\xi|\ov{m})$, so that Eq.~\eqref{eq:y_4} becomes \cite{Reichl_1984}
\begin{equation}
\ov{y}(\tau; u, {\sf e}) \;=\; 1 \;-\; s_{1}\;{\rm cn}\left(\xi\,|\, \ov{m}\right),
\label{eq:y_2}
\end{equation}
where $\xi = \ov{\nu}\tau$, with $\ov{\nu} = \sqrt{m}\,\nu = \sqrt{u{\sf e}}$, while the particle velocity in this case is
 \begin{equation}
 \ov{y}^{\prime}(\tau; u, {\sf e}) \;=\;  u\,(1 + {\sf e})\;\sqrt{m}\;{\rm sn}\left( \xi\,|\, \ov{m}\right)\,{\rm dn}\left( \xi\,|\, \ov{m}\right),
  \label{eq:y_II_velocity}
 \end{equation}
 and the period for this motion  (normalized to $\Omega_{0}$) is $\ov{T} = 4\,{\sf K}(\ov{m})/\ov{\nu} \equiv 4\ov{\sf K}/\ov{\nu}$. We can now calculate the orbit-averaged position $\langle \ov{y}\rangle \equiv \int_{0}^{\ov T} \ov y(\tau)\, d\tau/\ov T$ using Eq.~\eqref{eq:y_2}, which can be expressed as
\begin{equation}
\langle \ov y\rangle(u,{\sf e}) \;=\; 1 \;-\; \frac{\sqrt{\ov m}}{4\ov{\sf K}}\int_{0}^{4\ov{\sf K}} {\rm cn}(\xi|\ov m)\,d\xi \;=\; 1.
\end{equation}
Hence, no polarization shift is observed for a type II orbit, which is consistent with the breakdown of the guiding-center approximation since the particle orbit crosses the neutral sheet where the magnetic field vanishes.

\begin{figure}
\epsfysize=1.8in
\epsfbox{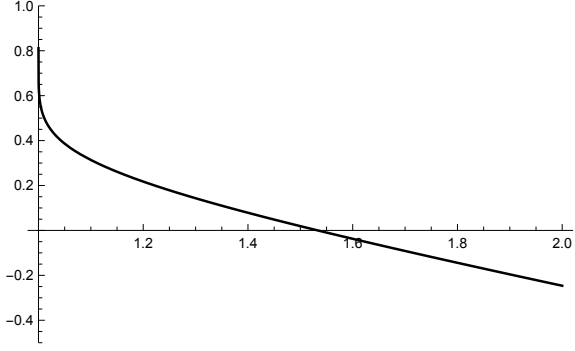}
\caption{Plot of the drift velocity $\langle \ov{x}^{\prime}\rangle/u$ in the range $1 \leq {\sf e} \leq 2$. Note that the drift velocity starts at $\langle \ov x^{\prime}\rangle(u,1) = u$ and experiences drift reversal when ${\sf E}(\ov m)/{\sf K}(\ov m) > 1/2$.}
\label{fig:ov_x_drift}
\end{figure}

From the solution \eqref{eq:y_2} for $\ov y(\tau; u, {\sf e})$, we can also obtain an expression for $\ov x^{\prime}(\tau; u, {\sf e})$ from Eq.~\eqref{eq:momentum}:
\begin{equation}
\ov x^{\prime}(\tau; u, {\sf e}) \;=\; u \left[ 1 \;-\frac{}{} (1+{\sf e})\,{\rm cn}^{2}(\ov{\nu}\,\tau|\ov m) \right].
\label{eq:x_II}
\end{equation}
When averaging the right side of Eq.~\eqref{eq:x_II} over an entire orbital period $\ov{T} = 4\ov{\sf K}/\ov\nu$, we find the orbit-averaged drift velocity
\begin{eqnarray}
\left\langle \ov x^{\prime}\right\rangle(u,{\sf e}) &=& u \left[ 1 \;-\; \frac{1+{\sf e}}{4\,\ov{\sf K}}\int_{0}^{4\ov{\sf K}} {\rm cn}^{2}(\xi)\,d\xi \right] \nonumber \\
 &=& u\,{\sf e} \left[ 1 \;-\; 2\,\frac{{\sf E}(\ov m)}{{\sf K}(\ov m)} \right],
\label{eq:ov_x_drift}
\end{eqnarray}
which is shown in Fig.~\ref{fig:ov_x_drift}, where a drift reversal is observed when ${\sf E}(\ov m)/{\sf K}(\ov m) > \frac{1}{2}$. A similar reversal is observed in the bounce-averaged toroidal drift precession frequency for the banana orbit of a trapped particle in an axisymmetric tokamak magnetic field \cite{Brizard_2011}. Next, the explicit integration of Eq.~\eqref{eq:x_II} yields the expression
\begin{eqnarray}
\ov{x}(\tau;u,{\sf e}) &=& u\,\tau \;-\; \frac{u}{\ov{\nu}}\,(1 + {\sf e}) \int_{0}^{\ov{\nu}\tau} {\rm cn}^{2}(\xi|\ov m)\,d\xi \nonumber \\
 &=&  u\,\tau \;-\;  2\,\ov{\nu} \left[ {\cal E}(\ov{\nu}\tau|\ov m) \;-\frac{}{} (1 - \ov{m})\,\ov{\nu}\,\tau\right] \nonumber \\
 &\equiv& \left\langle \ov{x}^{\prime}\right\rangle\,\tau \;-\; 2\,\ov{\nu}\;{\cal Z}(\ov{\nu}\tau,\ov m).
\end{eqnarray}
A plot of $\ov{x}(\tau;u,{\sf e})$ is shown in the bottom frame of Fig.~\ref{fig:x_drift} for $u = 1/2$ and ${\sf e} = 1.8$ in the range $0 \leq \tau \leq 2\ov{\sf K}/\ov{\nu}$, with the reversed drift motion $\langle \ov{x}^{\prime}\rangle\,\tau$ shown as a dashed line.

 \begin{figure}
\epsfysize=1.5in
\epsfbox{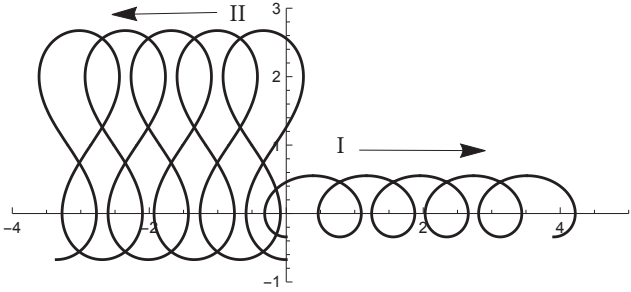}
\caption{Plots of particle orbits in the $(x,y)$-plane: (I) $y(\tau; u, {\sf e})$ versus $x(\tau; u, {\sf e})$ for $u = 1/2$ and ${\sf e} = 0.8$; and (II) $\ov y(\tau; u, {\sf e})$ versus $\ov x(\tau; u, {\sf e})$ for $u = 1/2$ and ${\sf e} = 1.8$. The drift motion along the $x$-axis for each orbit is shown by the arrow.}
\label{fig:xy_plots}
\end{figure}

\subsection{Separatrix Solution}

\begin{figure}
\epsfysize=1.7in
\epsfbox{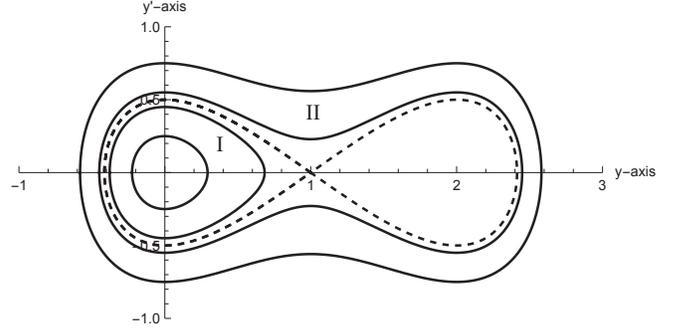}
\caption{Phase-space plots $(y,y^{\prime})$ for type I $(0 < {\sf e} < 1)$ and type II $({\sf e} > 1)$. In both cases, the parameter $u = 1/2$ is used and the separatrix $({\sf e} = 1)$ is shown as a dashed curve.}
\label{fig:phase_space}
\end{figure}

At the interface between these two orbit types is the separatrix solution for ${\sf e} = 1$ (i.e., $m = 1$):
\begin{equation}
y(\tau; u,1) \;=\; 1 \;-\; 2\,\sqrt{u}\;{\rm sech}\left(\sqrt{u}\,\tau\right),
\label{eq:y_S}
\end{equation}
which is recovered from both solutions \eqref{eq:y_4} and \eqref{eq:y_2}. The $x$-velocity, on the other hand, is
\begin{equation}
x^{\prime}(\tau; u, 1) \;=\; u \left[ 1 \;-\frac{}{} 2\;{\rm sech}^{2}(\sqrt{u}\,\tau) \right],
\end{equation}
which yields the $x$-position
\[ x(\tau; u, 1) \;=\; u\,\tau \;-\; 2\sqrt{u}\;\tanh(\sqrt{u}\,\tau). \]
Figure \ref{fig:phase_space} shows the phase-space plots $(y,y^{\prime})$ for type I and II for $u = 1/2$, with the separatrix solution $({\sf e} = 1)$ shown as a dashed curve.

\section{\label{sec:action}Action Integrals and Orbital Periods}

The area enclosed by each phase-space curve in Fig.~\ref{fig:phase_space} is related to the action integral $J(E) = (1/2\pi)\oint y^{\prime}\,dy$, from which the orbital period can be derived as $T(E) \equiv 2\pi\,dJ(E)/dE$. In our model, the energy is expressed as 
\begin{equation}
E \;=\; u^{2}{\sf e}^{2}/2, 
\label{eq:E_def}
\end{equation}
so that the orbital period $T(u,{\sf e})$ for each orbit type can be calculated as
\begin{equation}
T(u,{\sf e}) \;\equiv\; \frac{2\pi}{u^{2}{\sf e}}\;\pd{J(u,{\sf e})}{{\sf e}} \;\equiv\; \frac{2\pi}{\omega(u,{\sf e})},
\label{eq:T_JE}
\end{equation}
where $\omega(u,{\sf e})$ denotes the angular frequency for the orbit.

\subsection{Action Integrals}

The action integral for a type I orbit $({\sf e} < 1)$ is calculated from the solution \eqref{eq:y_4} as
\begin{eqnarray}
J &\equiv& \frac{1}{2\pi}\oint y^{\prime}\,dy \;=\; \frac{1}{2\pi}\int_{0}^{T} y^{\prime 2}\;d\tau \nonumber \\
 &=& \frac{2\,\nu^{3}}{\pi}\;m^{2} \int_{0}^{2\,{\sf K}} {\rm cn}^{2}\zeta\;{\rm sn}^{2}\zeta\;d\zeta,
 \end{eqnarray}
where $u\,(1 + {\sf e}) = 2\,\nu^{2}$ and $m = 2{\sf e}/(1 + {\sf e}) < 1$. Using standard properties of the the Jacobi elliptic functions [see Eq.~\eqref{eq:int_sn_cn} in App.~A], we find
 \begin{equation}
 J(u,{\sf e}) \;=\; \frac{4\,\nu^{3}}{3\pi} \left[ (2 - m)\,{\sf E}(m) \;-\frac{}{} 2\,(1 - m)\,{\sf K}(m) \right].
  \label{eq:J_I}
 \end{equation}
As ${\sf e} \rightarrow 1$ $(m \rightarrow 1)$, we find $J(u,1) = 4 u^{\frac{3}{2}}/(3\pi)$, while the action $J(u,{\sf e}) \sim (u/2)^{\frac{3}{2}} {\sf e}^{2}$ vanishes as ${\sf e} \rightarrow 0$.

 The action integral for a type II orbit (${\sf e} > 1$), on the other hand, is calculated from the solution \eqref{eq:y_2} as
\begin{eqnarray}
\ov{J} &\equiv& \frac{1}{2\pi}\oint \ov{y}^{\prime}\,dy \;=\; \frac{1}{2\pi}\int_{0}^{\ov{T}} \ov{y}^{\prime 2}\;d\tau \nonumber \\
 &=&  \frac{2\,\nu^{3}}{\pi}\;\sqrt{m} \int_{0}^{4\,\ov{\sf K}} {\rm sn}^{2}\xi\;{\rm dn}^{2}\xi\;d\xi.
 \end{eqnarray}
  Using standard properties of the the Jacobi elliptic functions  [see Eq.~\eqref{eq:int_sn_dn} in App.~A], we find
 \begin{equation}
 \ov{J}(u,{\sf e}) \;=\; \frac{8\,\nu^{3}}{3\pi}\;\sqrt{m} \left[ (2 - m)\,\ov{\sf E} \;-\frac{}{} (1-m)\;\ov{\sf K} \right].
 \label{eq:J_II}
 \end{equation}
As ${\sf e} \rightarrow 1$ $(m \rightarrow 1)$, we find $\ov{J}(u,1) = 8 u^{\frac{3}{2}}/(3\pi)$, which is double the limit for $J(u,1)$ since only one separatrix well is considered in the type I case (see Fig.~\ref{fig:phase_space}). On the other hand, as ${\sf e} \rightarrow \infty$ $(m \rightarrow 2)$, we find $\ov{J}(u,{\sf e}) \rightarrow 4\,(u\,{\sf e})^{\frac{3}{2}}{\sf K}(\frac{1}{2})/(3\pi)$.

\subsection{Orbital Periods}

For an orbit of type I, using Eq.~\eqref{eq:J_I}, we find
\begin{eqnarray}
T(u,{\sf e}) &=& \left(\frac{2\sqrt{2}\,u^{\frac{3}{2}}}{3\,u^{2}{\sf e}}\right) \left[ \frac{3\,{\sf e}}{\sqrt{1 + {\sf e}}}\;{\sf K}(m) \right] \nonumber \\
 &=& \frac{2\,{\sf K}(m)}{\sqrt{u\,(1 + {\sf e})/2}} \;\equiv\; \frac{2\,{\sf K}}{\nu} \;=\; \frac{2\pi}{\omega},
\end{eqnarray}
which is the orbital period observed from the type-I solution \eqref{eq:y_4}. For an orbit of type II, using Eq.~\eqref{eq:J_II}, we find
\begin{eqnarray}
\ov{T}(u,{\sf e}) &=& \left(\frac{4\sqrt{2}\,u^{\frac{3}{2}}}{3\,u^{2}{\sf e}}\right) \left[ 3\,\sqrt{\frac{{\sf e}}{2}}\;{\sf K}(\ov{m}) \right] \nonumber \\
 &=& \frac{4\,{\sf K}(\ov{m})}{\sqrt{u\,{\sf e}}} \;\equiv\; \frac{4\,\ov{\sf K}}{\ov{\nu}} \;=\; \frac{2\pi}{\ov\omega},
\end{eqnarray}
which is the orbital period observed from the type-II solution \eqref{eq:y_2}. 

\section{\label{sec:canonical}Angle Coordinates and Canonical Condition}

The canonical transformation from the canonical coordinates $(y,P_{y})$ to the action-angle coordinates $(\theta,J)$ satisfies the canonical condition 
\begin{equation}
\pd{y}{\theta}\,\pd{P_{y}}{J} \;-\; \pd{y}{J}\,\pd{P_{y}}{\theta} \;=\; 1,
\label{eq:canonical}
\end{equation}
where $y(\tau;u,{\sf e}) \equiv y[\theta(\tau;u,{\sf e}), J(u,{\sf e})]$ and $P_{y}(\tau;u,{\sf e}) \equiv \partial y(\tau;u,{\sf e})/\partial \tau = [\partial\theta(\tau;u,{\sf e})/\partial \tau]\,\partial y(\theta,J)/\partial\theta$. By definition, using the Hamiltonian $H = 
u^{2}{\sf e}^{2}/2$, the angle coordinate evolves according to the canonical Hamilton equation
\begin{equation}
\pd{\theta}{\tau} \;\equiv\; \pd{H}{J} \;=\; u^{2}{\sf e}\;\pd{\sf e}{J} \;=\; \omega(J),
\end{equation}
which was shown in Eq.~\eqref{eq:T_JE} to be correct for both orbital types. Since the Hamiltonian is independent of the canonical angle $\theta$, we obtain $\partial J/\partial\tau = -\,\partial H/\partial\theta \equiv 0$.

We note that the canonical transformation from the canonical coordinates $(y,P_{y})$ to the action-angle coordinates $(\theta,J)$ can be inverted, yielding the canonical condition
\begin{equation}
\pd{\theta}{y}\,\pd{J}{P_{y}} \;-\; \pd{\theta}{P_{y}}\,\pd{J}{y} \;=\; 1,
\label{eq:canonical_inverse}
\end{equation}
for the transformation from the action-angle coordinates $(\theta,J)$ to the canonical coordinates $(y,P_{y})$.

\subsection{Angle coordinates}

For type I orbits, we use the definition $\zeta(\theta,J) = \nu\tau \;=\; 2\,{\sf K}(m)\,\theta/(2\pi)$, which yields the angle coordinate
\begin{equation}
\theta \;=\; \pi\zeta/{\sf K}(m) \;\equiv\; \omega(u,{\sf e})\,\tau, 
\label{eq:theta_I}
\end{equation}
with $\partial\theta/\partial\tau = \pi\nu/{\sf K}(m) \equiv \omega(u,{\sf e})$. Hence,
\begin{equation}
\left. \begin{array}{rcl}
y(\theta,J) &=& 1 \;-\; 2\nu\;{\rm dn}\,\zeta(\theta,J)\\
 && \\
P_{y}(\theta,J) &=& 2u\,{\sf e}\,{\rm cn}\,\zeta(\theta,J)\;{\rm sn}\,\zeta(\theta,J)
\end{array} \right\},
\label{eq:yP_theta}
\end{equation}
where the action dependence appears through $J(u,{\sf e})$. 

For type II orbits, on the other hand, we use the definition $\xi(\ov{\theta},\ov{J}) = \ov{\nu}\tau \;=\; 4\,\ov{\sf K}\,\ov{\theta}/(2\pi)$, which yields the angle coordinate
\begin{equation}
\ov{\theta} \;=\; \pi\xi/(2\,\ov{\sf K}) \;\equiv\; \ov{\omega}(u,{\sf e})\,\tau, 
\label{eq:theta_II}
\end{equation}
with $\partial\ov{\theta}/\partial\tau = \pi\ov{\nu}/(2\ov{\sf K}) \equiv \ov{\omega}(u,{\sf e})$. Hence,
\begin{equation}
\left. \begin{array}{rcl}
\ov{y}(\ov{\theta},\ov{J}) &=& 1 \;-\; 2\nu\;{\rm cn}\,\xi(\ov{\theta},\ov{J})  \\
 && \\
\ov{P_{y}}(\ov{\theta},\ov{J}) &=& 2u\,{\sf e}\;\sqrt{\ov m}\,{\rm sn}\,\xi(\ov{\theta},\ov{J}) \;{\rm dn}\,\xi(\ov{\theta},\ov{J}) 
\end{array} \right\},
\label{eq:ov_yP_theta}
\end{equation}
where the action dependence appears through $\ov{J}(u,{\sf e})$. 

\subsection{Canonical condition}

\subsubsection{Type I orbit}

We now verify the canonical condition \eqref{eq:canonical} for orbit type I. First, we note that the derivatives with respect to the action coordinate must be calculated more carefully according to the expression
\begin{eqnarray} 
\left.\pd{}{J}\right|_{\theta} &=& \;\pd{\sf e}{J} \left( \left.\pd{}{\sf e}\right|_{\theta} \;+\; \frac{1}{\Omega_{\sf e}}\;\pd{}{\tau} \right) \nonumber \\
 &=& \frac{\omega(u,{\sf e})}{u^{2}\,{\sf e}}  \left( \left.\pd{}{\sf e}\right|_{\theta} \;+\; \frac{1}{\Omega_{\sf e}}\;\pd{}{\tau} \right),
\label{eq:partial_J}
\end{eqnarray}
where $\Omega_{\sf e}^{-1} \equiv \partial\tau/\partial{\sf e}|_{\theta}$, with 
\[ d\theta \;=\; (\pi\nu/{\sf K})\,d\tau \;+\; \pi\tau\,d{\sf e}\;\partial(\nu/{\sf K})/\partial{\sf e} \;=\; 0, \]
so that we find
\[  \frac{1}{\Omega_{\sf e}} \;=\; -\; \frac{\tau\,{\sf K}}{\nu} \pd{}{\sf e}\left( \frac{\nu}{\sf K}\right) \;=\; \frac{\tau}{2{\sf e}\,(1 - {\sf e})}\left( \frac{\sf E}{\sf K} - (1 - {\sf e}) \right). \]
Second, the angular derivatives can be evaluated as $\partial/\partial\theta = \omega^{-1}\,\partial/\partial\tau$, so that Eq.~\eqref{eq:canonical} becomes
\begin{eqnarray}
\pd{y}{\theta}\,\pd{P_{y}}{J} - \pd{y}{J}\,\pd{P_{y}}{\theta} &=& \omega^{-1}\pd{y}{\tau}\frac{\omega}{u^{2}{\sf e}} \left(\pd{P_{y}}{\sf e} + \frac{1}{\Omega_{\sf e}}\pd{P_{y}}{\tau}\right) \nonumber \\
 &&- \frac{\omega}{u^{2}{\sf e}} \left(\pd{y}{\sf e} + \frac{1}{\Omega_{\sf e}}\pd{y}{\tau}\right)\omega^{-1}\pd{P_{y}}{\tau} \nonumber \\
 &=& \pd{y}{\tau}\pd{P_{y}}{E} - \pd{y}{E}\pd{P_{y}}{\tau},
 \label{eq:canonical_I}
 \end{eqnarray}
where $\partial/\partial E = (u^{2}{\sf e})^{-1}\partial/\partial{\sf e}$. Here, using Eq.~\eqref{eq:yP_theta}, we find
 \begin{eqnarray*}
\pd{y}{\tau}\;\pd{P_{y}}{E} &=& 4\,{\rm sn}^{2}\zeta\;{\rm cn}^{2}\zeta + m\,(2 - m)\;\pd{}{m}\left({\rm sn}^{2}\zeta\frac{}{}{\rm cn}^{2}\zeta\right), \\
\pd{y}{E}\;\pd{P_{y}}{\tau} &=& \left({\rm cn}^{2}\zeta \;-\frac{}{} {\rm sn}^{2}\zeta\right) {\rm dn}^{2}\zeta \nonumber \\
  &&+ (2 - m)\left({\rm cn}^{2}\zeta \;-\frac{}{} {\rm sn}^{2}\zeta\right) \;\pd{\,{\rm dn}^{2}\zeta}{m},
 \end{eqnarray*}
 where we used ${\sf e} = m/(2 - m)$ and the $m$-derivatives of ${\rm pq}\,\zeta \equiv {\rm pq}(\zeta|m) = {\rm pq}({\sf K}(m)\theta/\pi|m)$ are evaluated at constant $\theta$. Next, using the identities ${\rm cn}^{2}\zeta + {\rm sn}^{2}\zeta = 1$ and 
 ${\rm dn}^{2}\zeta + m\, {\rm sn}^{2}\zeta = 1$, the canonical condition \eqref{eq:canonical_I} becomes
\begin{equation}
\pd{y}{\theta}\,\pd{P_{y}}{J} - \pd{y}{J}\,\pd{P_{y}}{\theta} \;=\; 1,
\end{equation}
where the final result is obtained by cancellations without evaluating the $m$-derivatives.

\subsubsection{Type II orbit}

Next, we verify the canonical condition \eqref{eq:canonical} for orbit type II. Here, we replace Eq.~\eqref{eq:partial_J} with the expression
\begin{eqnarray} 
\left.\pd{}{\ov J}\right|_{\theta} &=& \;\pd{\sf e}{\ov J} \left( \left.\pd{}{\sf e}\right|_{\ov\theta} \;+\; \frac{1}{\ov\Omega_{\sf e}}\;\pd{}{\tau} \right) \nonumber \\
 &=& \frac{\ov\omega(u,{\sf e})}{u^{2}\,{\sf e}}  \left( \left.\pd{}{\sf e}\right|_{\ov\theta} \;+\; \frac{1}{\ov\Omega_{\sf e}}\;\pd{}{\tau} \right),
\label{eq:partial_ovJ}
\end{eqnarray}
where $\ov\Omega_{\sf e}^{-1} \equiv \partial\tau/\partial{\sf e}|_{\ov\theta}$, with 
\[ d\ov\theta \;=\; (\pi\ov{\nu}/2\ov{\sf K})\,d\tau \;+\; (\pi\tau/2)\,d{\sf e}\;\partial(\ov{\nu}/\ov{\sf K})/\partial{\sf e} \;=\; 0, \]
so that we find
\[ \frac{1}{\ov\Omega_{\sf e}} \;=\; -\; \frac{\tau\,\ov{\sf K}}{\ov\nu} \pd{}{\sf e}\left( \frac{\ov\nu}{\ov{\sf K}}\right) \;=\; \frac{\tau}{2(1 - {\sf e}^{2})}\left( \frac{\ov{\sf E}}{\ov{\sf K}} - (1 - {\sf e}) \right). \]
Second, the angular derivatives can be evaluated as $\partial/\partial\ov\theta = \ov\omega^{-1}\,\partial/\partial\tau$, so that Eq.~\eqref{eq:canonical} becomes
\begin{eqnarray}
\pd{\ov y}{\ov\theta}\,\pd{\ov P_{y}}{\ov J} - \pd{\ov y}{\ov J}\,\pd{\ov P_{y}}{\ov\theta} &=& \ov\omega^{-1}\pd{\ov y}{\tau}\frac{\ov\omega}{u^{2}{\sf e}} \left(\pd{\ov P_{y}}{\sf e} + \frac{1}{\ov\Omega_{\sf e}}\pd{\ov P_{y}}{\tau}\right) \nonumber \\
 &&-\; \frac{\ov\omega}{u^{2}{\sf e}} \left(\pd{\ov y}{\sf e} + \frac{1}{\ov\Omega_{\sf e}}\pd{\ov y}{\tau}\right)\;\ov\omega^{-1}\pd{\ov P_{y}}{\tau} \nonumber \\
 &=& \pd{\ov y}{\tau}\pd{\ov P_{y}}{E} - \pd{\ov y}{E}\pd{\ov P_{y}}{\tau}.
  \label{eq:canonical_II}
 \end{eqnarray}
 Here, using Eq.~\eqref{eq:yP_theta}, we find
 \begin{eqnarray*}
\pd{\ov y}{\tau}\;\pd{\ov P_{y}}{E} &=&  (2 \ov m + 1)\,\ov{\rm sn}^{2}\xi\,\ov{\rm dn}^{2}\xi \nonumber \\
  &&-\; \ov m\,(2 \ov m - 1)\; \pd{}{\ov m}\left( \ov{\rm sn}^{2}\xi\frac{}{} \ov{\rm dn}^{2}\xi\right), \\
\pd{\ov y}{E}\;\pd{\ov P_{y}}{\tau} &=& \ov{\rm cn}^{2}\xi\;\left( \ov{\rm dn}^{2}\xi \;-\frac{}{} \ov m\;\ov{\rm sn}^{2}\xi \right) \nonumber \\
  &&+\; \ov m\,(2 \ov m - 1)\;\left( \ov{\rm dn}^{2}\xi \;-\frac{}{} \ov m\;\ov{\rm sn}^{2}\xi \right)\;\pd{\,\ov{\rm cn}^{2}\xi}{\ov m},
 \end{eqnarray*}
 where we used ${\sf e} = 1/(2 \ov m - 1)$ and the $\ov m$-derivatives of $\ov{\rm pq}\,\xi \equiv {\rm pq}(\xi|\ov m) = {\rm pq}(2{\sf K}(\ov m)\ov\theta/\pi| \ov m)$ are evaluated at constant $\ov\theta$. Next, using the identities $\ov{\rm cn}^{2}\xi + 
 \ov{\rm sn}^{2}\xi = 1$ and  $\ov{\rm dn}^{2}\xi + \ov m\, \ov{\rm sn}^{2}\xi = 1$, the canonical condition \eqref{eq:canonical_II} becomes
\begin{equation}
\pd{\ov y}{\ov\theta}\,\pd{\ov P_{y}}{\ov J} - \pd{\ov y}{\ov J}\,\pd{\ov P_{y}}{\ov\theta} \;=\; 1,
\end{equation}
where the final result is obtained by cancellations without evaluating the $\ov m$-derivatives.

\section{\label{sec:generating}Canonical Transformation}

The canonical condition \eqref{eq:canonical} proves the existence of a generating function $S$ for the canonical transformation \cite{Goldstein_2002} $(y,P_{y}) \rightarrow (\theta,J)$. Here, the periodic function $S$ is defined from the expression $P_{y}\;dy = J\;d\theta + dS$, which leads to  two complementary equations
\begin{equation}
\left. \begin{array}{rcl}
J \;+\; \partial S/\partial\theta &=& P_{y}\;\partial y/\partial\theta \\
 && \\
\partial S/\partial J &=& P_{y}\;\partial y/\partial J
\end{array} \right\},
\label{eq:canonical_eqs}
\end{equation}
which are consistent with the canonical condition \eqref{eq:canonical}. We note that the expression $P_{y}\,dy - dS = J\,d\theta$ is also consistent with the inverse canonical condition \eqref{eq:canonical_inverse}. 

In Eq.~\eqref{eq:canonical_eqs}, the partial derivative $\partial/\partial J$ is understood in the sense of Eq.~\eqref{eq:partial_J}. The first equation in Eq.~\eqref{eq:canonical_eqs} yields the action-integral definition
\[ J \;=\; \frac{1}{2\pi}\;\int_{0}^{2\pi} P_{y}\,\pd{y}{\theta}\;d\theta \;=\; \frac{1}{2\pi}\int_{0}^{T} \left(y^{\prime}\right)^{2} d\tau. \]
This equation also yields a defining expression for the generating function
\begin{eqnarray}
S(\theta,J) &\equiv& \int_{0}^{\theta/\omega} P_{y}\,\pd{y}{\tau}\;d\tau \;-\; J\,\theta \nonumber \\
 &=& \int_{0}^{\theta/\omega} \left(\pd{y}{\tau}\right)^{2}\;d\tau \;-\; J\,\theta,
\label{eq:S_def}
\end{eqnarray}
where $S(0,J) \equiv 0$ and we will show that $S(\theta,J)$ is explicitly dependent on the angle coordinate $\theta$, i.e., the orbit average $\langle S\rangle = 0$ vanishes for both orbit types.

\subsection{Type I Orbit}

For an orbit of type I, we use the orbit velocity \eqref{eq:y_I_velocity} and Eq.~\eqref{eq:S_def} yields
\begin{equation}
S(\theta,J) = 4\nu^{3}\,m^{2}\int_{0}^{\zeta}{\rm sn}^{2}(u|m)\;{\rm cn}^{2}(u|m)\,du \;-\; J\,\theta.
\end{equation}
Next, using the Jacobi zeta function \eqref{eq:Jacobi_zeta}, the integral can be evaluated as
\begin{eqnarray*} 
 &&4\nu^{3}\,m^{2}\int_{0}^{\zeta}{\rm sn}^{2}(u|m)\;{\rm cn}^{2}(u|m)\,du \\
 &=& \frac{4}{3}\,\nu^{3} \left[ (2 - m)\,{\cal Z}(\zeta,m) \;+\; \frac{1}{2}\pd{}{\zeta}{\rm dn}^{2}(\zeta|m) \right] \;+\; J\,\theta,
 \end{eqnarray*}
and the canonical generating function for orbit type I is
\begin{eqnarray}
S(\theta,J) &=& \frac{4}{3}\,\nu^{3} \left[ (2 - m)\,{\cal Z}(\zeta,m) \;+\; \frac{1}{2}\pd{}{\zeta}{\rm dn}^{2}(\zeta|m) \right] \nonumber \\
 &\equiv&  \frac{2}{3}\,u\,(1 + {\sf e})\,\omega(u,{\sf e})\;\pd{R(\theta,J)}{\theta},
\label{eq:S_I}
\end{eqnarray}
where the periodic function
\begin{eqnarray} 
R(\theta,J) &\equiv& (2 - m)\,\ln\vartheta_{4}(\theta/2,{\sf q}(m)) \nonumber \\
 &&+\; \frac{1}{2}\,(1 - m)^{\frac{1}{2}}\;\frac{\vartheta_{3}^{2}(\theta/2,{\sf q}(m))}{\vartheta_{4}^{2}(\theta/2,{\sf q}(m))}
 \end{eqnarray}
is expressed in terms of the theta functions \cite{Lawden_1989,NIST_Jacobi}
\begin{equation}
\left. \begin{array}{rcl}
\vartheta_{3}(u,{\sf q}) &=& 1 + 2\,\sum_{n = 1}^{\infty} {\sf q}^{n^{2}}\,\cos(2 n u) \\
 && \\
\vartheta_{4}(u,{\sf q}) &=& 1 + 2\,\sum_{n = 1}^{\infty} (-1)^{n} {\sf q}^{n^{2}}\,\cos(2n u)
\end{array} \right\},
\end{equation}
which are defined in terms of the nome 
\[ {\sf q}(m) \;=\; \exp[-\pi{\sf K}(1-m)/{\sf K}(m)]. \]
We note that Eq.~\eqref{eq:S_I} automatically yields the vanishing orbit average $\langle S\rangle = 0$.

\subsection{Type II Orbit}

For an orbit of type II, we use the orbit velocity \eqref{eq:y_II_velocity} and Eq.~\eqref{eq:S_def} yields
\begin{equation}
\ov S(\ov\theta, \ov J) = 4\nu^{3}\,\sqrt{m}\int_{0}^{\xi}{\rm sn}^{2}(u|\ov m)\;{\rm dn}^{2}(u|\ov m)\,du - \ov J\,\ov\theta.
\end{equation}
Next, using the Jacobi zeta function \eqref{eq:Jacobi_xi}, the integral can be evaluated as
\begin{eqnarray*}
&&4\nu^{3}\,\sqrt{m}\int_{0}^{\xi}{\rm sn}^{2}(u|\ov m)\;{\rm dn}^{2}(u|\ov m)\,du \\
 &=& \frac{4}{3}\,\nu^{3}\,\sqrt{m} \left[ (2 - m)\,{\cal Z}(\xi,\ov m) \;+\; \frac{1}{2}\pd{}{\xi}{\rm cn}^{2}(\xi|\ov m) \right] \;+\; \ov J\,\ov\theta,
 \end{eqnarray*}
and the canonical generating function for orbit type II is
\begin{eqnarray}
\ov S(\ov\theta, \ov J) &=& \frac{4}{3}\,\nu^{3}\,\sqrt{m} \left[ (2 - m)\,{\cal Z}(\xi,\ov m) \;+\; \frac{1}{2}\pd{}{\xi}{\rm cn}^{2}(\xi|\ov m) \right] \nonumber \\
 &\equiv&  \frac{1}{3}\,u\,(1 + {\sf e})\,\ov\omega(u,{\sf e})\;\pd{\ov R(\ov\theta, \ov J)}{\ov\theta},
\label{eq:S_II}
\end{eqnarray}
where the periodic function
\begin{eqnarray} 
\ov R(\ov\theta, \ov J) &\equiv& (2 - m)\,\ln\vartheta_{4}(\ov\theta,{\sf q}(\ov m)) \nonumber \\
 &&+\; \frac{1}{2}(m - 1)^{\frac{1}{2}}\;\frac{\vartheta_{2}^{2}(\ov\theta,{\sf q}(\ov m))}{\vartheta_{4}^{2}(\ov\theta,{\sf q}(\ov m))}
 \end{eqnarray}
is expressed in terms of the theta function $\vartheta_{4}(\ov\theta,{\sf q}(\ov m))$ and the theta function 
\begin{equation}
\vartheta_{2}(\ov\theta,\ov{\sf q}) \;=\; 2\,\sum_{n = 0}^{\infty} \ov{\sf q}^{(n + \frac{1}{2})^{2}}\,\cos[(2n +1)\,\ov\theta],
\end{equation}
which are defined in terms of the nome 
\[ \ov{\sf q} \;=\; {\sf q}(\ov m) \;=\; \exp[-\pi{\sf K}(1-\ov m)/{\sf K}(\ov m)]. \] 
Once again, we note that Eq.~\eqref{eq:S_II} automatically yields the vanishing orbit average  $\langle S\rangle = 0$.

\section{\label{sec:electric}Motion in the presence of an electric field}

In this Section, we consider the impact of a constant electric field on the dynamics of a charged particle in a straight magnetic field with constant perpendicular gradient. The case of a constant electric field ${\bf E} = E_{0}\,\wh{\sf x}$ in the ignorable 
$x$-direction has been studied in Ref.~\cite{Parks_2003} and will not be considered here. 

Instead, we consider the constant electric field ${\bf E} = E_{0}\,\wh{\sf y}$, so that the equations of motion \eqref{eq:xy_exact} become 
\begin{equation}
\left. \begin{array}{rc l}
\ddot{x} &=& \Omega_{0}\,(1 - y/L)\,\dot{y} \\
\ddot{y} &=& qE_{0}/M \;-\; \Omega_{0}\,(1 - y/L)\,\dot{x}
\end{array} \right\}.
\label{eq:xy_E}
\end{equation}
These equations still conserve the canonical momentum \eqref{eq:canonical_momentum} while the energy conservation law
\[ E \;=\; M(\dot{x}^{2} + \dot{y}^{2})/2 - qE_{0}\,(y - 1) \]
is now expressed in terms of an electrostatic potential that is assumed to vanish at $y = 1$. The normalized equation of motion \eqref{eq:y_pp} is now replaced with
\begin{equation}
y^{\prime\prime} \;=\; w \;-\; (1 - y)\,x^{\prime} \;=\; w \;-\; (1 - y)\,u \;+\; \frac{1}{2}\,(1 - y)^{3},
\label{eq:y_pp_w}
\end{equation}
where $w \equiv qE_{0}/(M\Omega_{0}^{2}L)$ is the normalized $E \times B$ velocity. 

\subsection{Asymmetric quartic potential}

\begin{figure}
\epsfysize=1.6in
\epsfbox{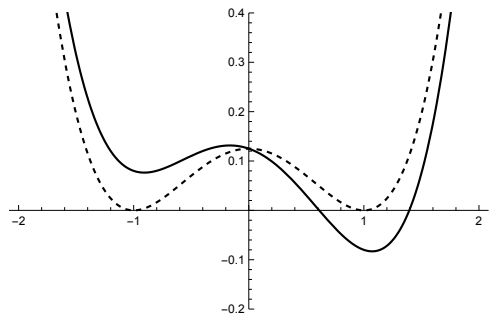}
\caption{Plot of the asymmetric quartic potential \eqref{eq:V_yw} in the range $-2 \leq s \leq 2$ for $u = 1/2$ and $w = -0.08$ (solid). The symmetric quartic potential, with $w = 0$, is shown as a dashed curve.}
\label{fig:V_yw}
\end{figure}

We now set $s = 1 - y$, so that Eq.~\eqref{eq:y_pp_w} becomes
\begin{equation}
s^{\prime\prime} \;=\; -\,w \;+\; u\,s \;-\; \frac{1}{2}\,s^{3} \;\equiv\; -\,V^{\prime}(s),
\end{equation}
where the asymmetric quartic potential
\begin{eqnarray}
V(s) &=& w\,s \;+\; \frac{1}{2} \left( u - \frac{1}{2}\,s^{2}\right)^{2} \nonumber \\
 &=& \frac{1}{2}\,u^{2} \;+\; w\,s \;-\; \frac{1}{2}\,u\,s^{2} \;+\; \frac{1}{8}\,s^{4}
\label{eq:V_yw}
\end{eqnarray}
has extrema points $(s_{1},s_{2},s_{3})$ at
\begin{equation}
\left. \begin{array}{rcl}
s_{1} &=& 2\,\sigma\;\cos(\pi/6 + \psi/3) \\
s_{2} &=& 2\,\sigma\;\sin(\psi/3) \\
s_{3} &=& -\,2\,\sigma\;\cos(\pi/6 - \psi/3)
\end{array} \right\},
\end{equation}
where $\sigma^{2} = 2\,u/3$ and $\psi = \sin^{-1}(w/\sigma^{3})$. These points are real when $-\sigma^{3} \leq w \leq \sigma^{3}$, so that the quartic potential \eqref{eq:V_yw} has two minima $(s_{1},s_{3})$ and a local maximum $(s_{2})$, where $s_{3} < s_{2} < s_{1}$. As shown in Fig.~\ref{fig:V_yw} (where $-\sigma^{3} < w < 0$), a particle can now be trapped either in a shallow well (left) or a deep well (right) if the energy level falls below the barrier height $V_{2} = V(s_{2})$. 

The problem of a particle moving in the asymmetric quartic potential \eqref{eq:V_yw} was solved exactly in terms of Jacobi and Weierstrass elliptic functions in Ref.~\cite{Brizard_Westland_2017}. There, it was shown that a particle trapped in the shallow well at a given energy level has exactly the same orbital period as a particle trapped in the deep well at the same energy level. While it should be possible to construct action-angle coordinates for the trapped motion in the shallow well and the deep well separately, as well as for the trapped motion above the potential barrier, we will not pursue this avenue in the present paper.

\subsection{Electric field perturbation}

In previous work, Reichl and Zheng \cite{Reichl_1984} considered a symmetric double-well potential perturbed by a small time-dependent (monochromatic) electric field represented in dimensionless form as $w = \epsilon\,\cos(\Omega\, \tau)$, which allows a fraction of particles trapped in one well to migrate over the top of the potential barrier. The analysis of this perturbed Duffing equation can also be found in Ref.~\cite{Nayfeh_1973}.

In this Section, we consider the case of a weak time-independent electric field $w = \epsilon\,\sigma^{3}$, with $\epsilon \ll 1$, so that the asymmetry in the quartic potential is weak. This perturbed dynamical problem can be represented as a Hamiltonian perturbation problem, where, using the action-angle coordinates $(J,\theta)$, the Hamiltonian is expressed as
\begin{equation}
H(J,\theta) \;=\; H_{0}(J) \;+\; \epsilon\,\sigma^{3}\,s(J,\theta),
\end{equation}
where $H_{0}(J) = \frac{1}{2}\,u^{2}\,{\sf e}^{2}(J)$ and $s(J,\theta)$ describes a solution of the unperturbed (symmetric) quartic potential problem. Because of the electric-field perturbation, the unperturbed action coordinate $J$ is no longer an invariant
\begin{equation}
\pd{J}{\tau} \;=\; -\;\pd{H}{\theta}(J,\theta) \;=\; -\; \epsilon\,\sigma^{3}\,\pd{s}{\theta}(J,\theta),
\end{equation}
while the angle coordinate evolves according the Hamilton equation
\begin{equation}
\pd{\theta}{\tau} \;=\; \pd{H}{J}(J,\theta) \;=\; \omega_{0}(J) \;+\; \epsilon\,\sigma^{3}\,\pd{s}{J}(J,\theta),
\end{equation}
where $\omega_{0}(J) = u^{2}{\sf e}\,\partial{\sf e}/\partial J$ is obtained from Eq.~\eqref{eq:T_JE}.

According to standard perturbation methods \cite{Nayfeh_1973}, we now seek a canonical transformation $(J,\theta) \rightarrow ({\cal J},\Theta)$, where the new action-angle coordinates $({\cal J},\Theta)$ are constructed by Lie-transform methods (see App.~C of Ref.~\cite{Brizard_2015})
\begin{eqnarray}
{\cal J} &=& J \;+\; \epsilon\,\{S_{1}, J\} \;+\;  \epsilon^{2} \{S_{2}, J\} \nonumber \\
 &&+\; \frac{\epsilon^{2}}{2}\left\{ S_{1},\frac{}{}\{S_{1}, J\} \right\} + \cdots, \label{eq:J_j} \\
\Theta &=& \theta \;+\; \epsilon\,\{S_{1}, \theta\} \;+\; \epsilon^{2} \{S_{2}, \theta\} \nonumber \\
 &&+\; \frac{\epsilon^{2}}{2}\left\{ S_{1},\frac{}{}\{S_{1}, \theta\} \right\} + \cdots. \label{eq:Theta_theta}
\end{eqnarray}
Here, the generating functions $S_{1}(J,\theta)$ and $S_{2}(J,\theta)$ are used to remove the angular dependence from the new Hamiltonian up to second order
\begin{eqnarray}
{\cal H}(J) &\equiv& H(J,\theta) \;-\; \epsilon\,\{S_{1}, H\} \;-\; \epsilon^{2}\,\{ S_{2}, H\} \nonumber \\
 &&+\; \frac{\epsilon^{2}}{2}\;\left\{ S_{1},\frac{}{} \{S_{1}, H\}\right\} \;+\; \cdots \nonumber \\ 
 &=& {\cal H}_{0}(J) \;+\; \epsilon\,{\cal H}_{1}(J) \;+\; \epsilon^{2}{\cal H}_{2}(J) + \cdots,
\label{eq:H_J}
\end{eqnarray}
and the action-angle Poisson bracket $\{\;,\;\}$ is defined as
\[
\{ f,\; g\} = \pd{f}{\theta}\;\pd{g}{J} - \pd{f}{J}\;\pd{g}{\theta} \equiv \pd{}{J}\left(g\;\pd{f}{\theta}\right) - \pd{}{\theta}\left(g\;\pd{f}{J}\right).
\]
By definition, the zeroth-order Hamiltonian is ${\cal H}_{0}(J) \equiv H_{0}(J)$, and the lowest-order (unperturbed) angular velocity is $\omega_{0}(J) \equiv {\cal H}_{0}^{\prime}(J)$.

From Eq.~\eqref{eq:H_J}, the new first-order Hamiltonian is defined by the relation
\begin{equation}
{\cal H}_{1}(J) \;=\; \sigma^{3}\,s(J,\theta) \;-\; \omega_{0}(J)\;\pd{S_{1}}{\theta},
\label{eq:H1_eq}
\end{equation}
and, since the left side of Eq.~\eqref{eq:H1_eq} is independent of the angle $\theta$, we readily find the solution for the new first-order Hamiltonian 
\begin{equation} 
{\cal H}_{1}(J) \;\equiv\; \sigma^{3}\,\langle s\rangle \;=\; \frac{\sigma^{3}}{2\pi}\int_{0}^{2\pi}s(J,\theta)\,d\theta,
\label{eq:H1_def}
\end{equation}
while the first-order generating function is defined as the non-secular solution $S_{1}(J,\theta)$ obtained from the angle-dependent part of Eq.~\eqref{eq:H1_eq}:
\begin{equation}
\pd{S_{1}}{\theta} \;=\; \frac{\sigma^{3}}{\omega_{0}}\left( s \;-\frac{}{} \langle s\rangle\right) \;\equiv\; \frac{\sigma^{3}}{\omega_{0}}\,\wt{s}.
\label{eq:S1_eq}
\end{equation}
At second order, the new Hamiltonian is defined from the expression
\begin{equation}
{\cal H}_{2}(J) = -\;\omega_{0}\;\pd{S_{2}}{\theta} - \omega_{1}\;\pd{S_{1}}{\theta} - \frac{\sigma^{3}}{2}\left\{ S_{1},\;\wt{s}(J,\theta)\right\},
\label{eq:H2_def}
\end{equation}
where $\omega_{1}(J) \equiv {\cal H}_{1}^{\prime}(J)$. Since the left side of Eq.~\eqref{eq:H2_def} is $\theta$-independent, the new second-order Hamiltonian is then defined as
\begin{eqnarray}
{\cal H}_{2}(J) &=& -\; \frac{\sigma^{3}}{2}\left\langle\left\{ S_{1},\frac{}{} s(J,\theta)\right\}\right\rangle \equiv -\; \frac{\sigma^{3}}{2}\pd{}{J}\left\langle \wt{s}\;\pd{S_{1}}{\theta}\right\rangle \nonumber \\
 &=& -\,\frac{\sigma^{6}}{2}\,\pd{}{J}\left[\frac{1}{\omega_{0}}\left( \langle s^{2}\rangle \;-\frac{}{} \langle s\rangle^{2} \right) \right].
 \label{eq:Ham_2nd}
\end{eqnarray}
The second-order generating function is defined as the non-secular solution $S_{2}(J,\theta)$ obtained from the angle-dependent part of Eq.~\eqref{eq:H2_def}.

Once the new Hamiltonian \eqref{eq:H_J} is constructed up to any desired order in $\epsilon$ (say $\epsilon^{n}$), the new action ${\cal J}$ is conserved up the $n$th-order, i.e., $d{\cal J}/d\tau = {\cal O}(\epsilon^{n+1})$, and the angular motion $d\Theta/d\tau = {\cal H}^{\prime}({\cal J}) \equiv \omega({\cal J})$ is exactly solved as $\Theta(\tau,{\cal J}) =  \omega({\cal J})\,\tau$. 

\subsubsection{Type I orbit}

We begin our perturbation analysis of the particle motion in an asymmetric quartic potential with type-I particle orbit, where $s(J,\theta) = 2\nu\,{\rm dn}(\theta\,{\sf K}/\pi|m)$. Using the definition for the quasi-periodic Jacobi amplitude function \cite{NIST_Jacobi}
\[ \int_{0}^{\zeta}{\rm dn}(u|m)\,du \;=\; \phi(\zeta|m) \;\equiv\; \sin^{-1}\left({\rm sn}(\zeta|m)\right), \]
with $\phi(\zeta + 2{\sf K}|m) = \phi(\zeta|m) + \pi$, we find the orbit average
\[ \langle s\rangle \;=\; \frac{\nu}{\sf K}\int_{0}^{2{\sf K}}{\rm dn}\,\zeta\; d\zeta \;=\; \frac{\pi\nu}{\sf K} \;\equiv\; \omega_{0}(J), \]
so that the new first-order Hamiltonian \eqref{eq:H1_def} becomes
\begin{equation}
{\cal H}_{1}(J) \;=\; \sigma^{3}\,\omega_{0}(J).
\label{eq:H1_sn}
\end{equation}
The generating function $S_{1}$ can be obtained from Eq.~\eqref{eq:H1_eq} by direct integration
\begin{eqnarray}
S_{1}(J,\theta) &=& \frac{2\nu\,\sigma^{3}}{\omega_{0}} \left( \int {\rm dn}(\theta\,{\sf K}/\pi|m)\,d\theta \;-\; \frac{\pi\theta}{2{\sf K}}\right) \nonumber \\
 &=& \sigma^{3} \left[ 2\,\phi(\theta\,{\sf K}/\pi|m) \;-\frac{}{} \theta \right],
 \label{eq:S1_sol}
\end{eqnarray}
which can be used in the first-order canonical transformation in Eqs.~\eqref{eq:J_j}-\eqref{eq:Theta_theta}.

Next, we calculate the new second-order Hamiltonian from Eq.~\eqref{eq:Ham_2nd}, where $\langle s\rangle = \omega_{0}$ and
\[ \langle s^{2}\rangle \;=\; \frac{2\nu^{2}}{\sf K}\int_{0}^{2{\sf K}}{\rm dn}^{2}\zeta\,d\zeta \;=\; 4\nu^{2}\,{\sf E}/{\sf K} = (4\nu/\pi)\,\omega_{0}{\sf E}, \]
so that Eq.~\eqref{eq:Ham_2nd} yields
\begin{equation}
{\cal H}_{2}(J) \;=\; -\,\frac{2\,\sigma^{6}}{\pi}\,\pd{}{J}\left( \nu\,{\sf E} \;-\; \frac{\pi}{4}\;\omega_{0}\right),
\label{eq:H2_I}
\end{equation} 
which yields the new Hamiltonian \eqref{eq:H_J} up to second order.

\subsubsection{Type II orbit}

Next, we consider the perturbation analysis of the particle motion in an asymmetric quartic potential with type-II particle orbits, where $s(\ov J,\ov\theta) = 2\nu\,{\rm cn}(2\ov\theta\,\ov{\sf K}/\pi|\ov m)$. Using the integral
\[ \int_{0}^{\xi}{\rm cn}(u|\ov m)\,du \;=\; \sqrt{m}\;\tan^{-1}\left(\sqrt{\ov m}\frac{}{}{\rm sd}(\xi|\ov m) \right), \]
we find $\langle s\rangle = 0$ and the new first-order Hamiltonian is ${\cal H}_{1}(\ov{J}) = 0$. The first-order generating function is thus defined by the relation
\begin{equation}
\pd{S_{1}}{\ov\theta} \;=\; \frac{2\nu\sigma^{3}}{\ov{\omega}_{0}}\;{\rm cn}(2\ov\theta\,\ov{\sf K}/\pi|\ov m),
\end{equation}
which yields the solution
\begin{eqnarray}
S_{1} &=& \frac{2\nu\sigma^{3}}{\ov{\omega}_{0}}\int_{0}^{\ov\theta}{\rm cn}(2\theta\,\ov{\sf K}/\pi|\ov m)\;d\theta = \frac{\pi\nu\sigma^{3}}{\ov{\sf K}\ov{\omega}_{0}}\int_{0}^{\xi} {\rm cn}(u|\ov{m})\,du \nonumber \\
 &=& 2\,\sigma^{3}\tan^{-1}\left(\sqrt{\ov m}\frac{}{}{\rm sd}(\xi|\ov m) \right),
 \label{eq:ov_S1_sol}
\end{eqnarray}
which can be used in the first-order canonical transformation in Eqs.~\eqref{eq:J_j}-\eqref{eq:Theta_theta}.

Next, we calculate the new second-order Hamiltonian from Eq.~\eqref{eq:Ham_2nd}, where
\[ \langle s^{2}\rangle \;=\; \frac{\nu^{2}}{\ov{\sf K}}\int_{0}^{4\ov{\sf K}}{\rm cn}^{2}\xi\,d\xi \;=\; \frac{4m\nu^{2}}{\ov{\sf K}}\,\left[ \ov{\sf E} \;-\frac{}{} (1 - \ov m)\,\ov{\sf K}\right], \]
so that Eq.~\eqref{eq:Ham_2nd} yields
\begin{equation}
{\cal H}_{2}(\ov J) \;=\; -\,\frac{4\,\sigma^{6}}{\pi}\,\pd{}{\ov J}\left[ \ov{\nu}\,\left(  \ov{\sf E} \;-\frac{}{} (1 - \ov m)\,\ov{\sf K}\right) \right],
\label{eq:H2_I}
\end{equation} 
which yields the new Hamiltonian \eqref{eq:H_J} up to second order.

\subsubsection{Approximate particle orbit}

The next step in the perturbative solution of the equations of motion \eqref{eq:xy_E} is to invert the action-angle transformations \eqref{eq:J_j}-\eqref{eq:Theta_theta}
\begin{eqnarray}
J(\tau,{\cal J}) &=& {\cal J} \;-\; \epsilon\,\{S_{1}, {\cal J}\} \;-\;  \epsilon^{2} \{S_{2}, {\cal J}\} \nonumber \\
 &&+\; \frac{\epsilon^{2}}{2}\left\{ S_{1},\frac{}{}\{S_{1}, {\cal J}\} \right\} + \cdots, \label{eq:j_J} \\
\theta(\tau,{\cal J}) &=& \Theta \;-\; \epsilon\,\{S_{1}, \Theta\} \;-\; \epsilon^{2} \{S_{2}, \Theta\} \nonumber \\
 &&+\; \frac{\epsilon^{2}}{2}\left\{ S_{1},\frac{}{}\{S_{1}, \Theta\} \right\} + \cdots, \label{eq:theta_Theta}
\end{eqnarray}
where $\Theta(\tau,{\cal J}) =  \omega({\cal J})\,\tau$, with $\omega({\cal J}) \equiv {\cal H}^{\prime}({\cal J})$, is used after derivatives with respect to $\Theta$ have been taken. Lastly, we insert Eqs.~\eqref{eq:j_J}-\eqref{eq:theta_Theta} into either the solution $y(J,\theta)$ for type I or II orbits, given by Eq.~\eqref{eq:yP_theta} or \eqref{eq:ov_yP_theta}, respectively. This perturbative solution was shown in Ref.~\cite{Brizard_2015} to be valid up to time scales of order $\epsilon^{-1}$ for the case of the perturbed harmonic oscillator.

\section{\label{sec:summary}Summary}

In this paper, we presented in Sec.~\ref{sec:orbits} all orbital solutions for the equations of motion \eqref{eq:xy_exact} describing the motion of a charged particle in a straight magnetic field with constant gradient. In addition, we derived expressions for the action-angle coordinates for each orbit type in Secs.~\ref{sec:action}-\ref{sec:canonical} and obtained an explicit expression for the generating function for each canonical transformation in Sec.~\ref{sec:generating}. This work follows in the steps of previous work on the trapped and passing particle orbits in simple tokamak geometry \cite{Brizard_Duthoit_2014}, based on the canonical transformation for the pendulum problem \cite{Brizard_2013}.

Next, we considered the case of motion in an asymmetric quartic potential due to the presence of a constant electric field in Sec.~\ref{sec:electric}, which is solved perturbatively for the case of a weak electric field. In future work, we will construct the canonical transformation to action-angle coordinates based on the exact solution in terms of elliptic functions presented in Ref.~\cite{Brizard_Westland_2017}.

\appendix

\section{\label{sec:App}Integrals of Jacobi Elliptic Functions}

In this Appendix, we present a brief review of the Jacobi epsilon and zeta functions \cite{Lawden_1989,NIST_Jacobi}.

\subsection{Jacobi Epsilon Function}

In calculating action integrals in Sec.~\ref{sec:action}, we need to compute integrals of even powers of ${\rm sn}(u|m)$: 
\[ I_{2n}(z|m) \;\equiv\; \int_{0}^{z}{\rm sn}^{2n}(u|m)\,du \]
for $n \geq 0$. First, $I_{0}(z|m) = z$ and
\begin{equation}
I_{2}(z|m) \;=\; \int_{0}^{z}{\rm sn}^{2}(u|m) du \;=\; m^{-1} \left[ z \;-\; {\cal E}(z|m)\right],
\end{equation}
where ${\cal E}(z|m)$ is the Jacobi epsilon function, which is quasi-periodic: ${\cal E}(z + 2{\sf K}|m) = {\cal E}(z|m) + 2\,{\sf E}(m)$. Next, we use the recurrence relation $(n \geq 1)$ \cite{Lawden_1989}
\begin{eqnarray}
m\,(1 + 2n)\,I_{2n+2}(z) &=& (1 + m)\,2n\;I_{2n}(z) \label{eq:recurrence} \\
 &&-\; (2n - 1)\,I_{2n-2}(z) \nonumber \\
 &&+\; {\rm sn}^{2n-1}(z)\,{\rm cn}(z)\,{\rm dn}(z), \nonumber
\end{eqnarray}
which, for $n = 1$, yields 
\begin{eqnarray*} 
3\,m\,I_{4}(z|m) &=& 2(1 + m)\,I_{2}(z|m) \;-\; I_{0}(z|m) \\
 &&+\; {\rm sn}(z|m)\,{\rm cn}(z|m)\,{\rm dn}(z|m),
 \end{eqnarray*}
or
\begin{eqnarray}
m\int_{0}^{z}{\rm sn}^{4}(u|m) du &=& \frac{1}{3}\,\left( {\rm sn}(z|m)\,{\rm cn}(z|m)\,{\rm dn}(z|m) \;-\frac{}{} z \right) \nonumber \\
 &&+\; \frac{2}{3m}(1 + m) \left(z \;-\frac{}{} {\cal E}(z|m)\right).
\end{eqnarray}

First, using the identity ${\rm cn}^{2}(z|m) = 1 - {\rm sn}^{2}(z|m)$, we find
\begin{eqnarray}
m^{2}\int_{0}^{\zeta}{\rm sn}^{2}u\;{\rm cn}^{2}u\,du &=& m^{2}\int_{0}^{\zeta}\left( {\rm sn}^{2}u \;-\frac{}{} {\rm sn}^{4}u \right)du \nonumber \\
 &=& m^{2} \left[ I_{2}(\zeta|m) \;-\frac{}{} I_{4}(\zeta|m) \right] \nonumber \\
 &=& m \left(\zeta \;-\frac{}{} {\cal E}(\zeta)\right) \nonumber \\
  &&-\; \frac{m}{3}\,\left( {\rm sn}\,\zeta\;{\rm cn}\,\zeta\;{\rm dn}\,\zeta \;-\frac{}{} \zeta \right) \nonumber \\
   &&-\; \frac{2}{3}\,(1 + m) \left(\zeta \;-\frac{}{} {\cal E}(\zeta)\right) \nonumber \\
  &=& \frac{1}{3} \left[ (2 - m)\,{\cal E}(\zeta) \;-\frac{}{} 2\,(1 - m)\,\zeta \right] \nonumber \\
   &&-\; \frac{m}{3}\;{\rm sn}\,\zeta\;{\rm cn}\,\zeta\;{\rm dn}\,\zeta.
  \label{eq:int_sn_cn_zeta}
\end{eqnarray}
When evaluated at $\zeta = 2\,{\sf K}(m)$, this expression yields
\begin{eqnarray}
m^{2}\int_{0}^{2{\sf K}}{\rm sn}^{2}u\,{\rm cn}^{2}u\,du &=& \frac{2}{3}  (2 - m)\,{\sf E}(m) \nonumber \\
 &&-\; \frac{4}{3}\,(1 - m)\,{\sf K}(m),
\label{eq:int_sn_cn}
\end{eqnarray}
where we used ${\cal E}(2{\sf K}) = 2\,{\sf E}$. 

Second, using the identity ${\rm dn}^{2}(\xi|\ov{m}) = 1 - \ov{m}\,{\rm sn}^{2}(\xi|\ov{m})$, we find
\begin{eqnarray}
\int_{0}^{\xi}\ov{\rm sn}^{2}u\;\ov{\rm dn}^{2}u\,du &=& \int_{0}^{\xi}\left( \ov{\rm sn}^{2}u \;-\; \ov{m}\int_{0}^{\xi}\ov{\rm sn}^{4}u \right) du \nonumber \\
 &=& I_{2}(\xi| \ov m) \;-\; \ov{m}\;I_{4}(\xi| \ov m) \nonumber \\
 &=& m \left(\xi \;-\frac{}{} \ov{\cal E}(\xi)\right) \nonumber \\
  &&- \frac{1}{3} \left( \ov{\rm sn}\,\xi\;\ov{\rm cn}\,\xi\;\ov{\rm dn}\,\xi \;-\frac{}{} \xi \right) \nonumber \\
   &&- \frac{2}{3}(1 + m) \left(\xi \;-\frac{}{} \ov{\cal E}(\xi)\right) \nonumber \\
  &=& \frac{1}{3} \left[ (2 - m)\,\ov{\cal E}(\xi) \;-\frac{}{}  (1 - m)\,\xi \right] \nonumber \\
   &&-\; \frac{1}{3}\;\ov{\rm sn}\,\xi\;\ov{\rm cn}\,\xi\;\ov{\rm dn}\,\xi,
  \label{eq:int_sn_dn_xi}
\end{eqnarray}
where $\ov{m} \equiv m^{-1}$. When evaluated at $\xi = 4\,\ov{\sf K}$, this expression yields
\begin{equation}
\int_{0}^{4\ov{\sf K}}\ov{\rm sn}^{2}u\;\ov{\rm dn}^{2}u\,du \;=\; \frac{4}{3} \left[ (2 - m)\,\ov{\sf E} \;-\frac{}{} (1 - m)\,\ov{\sf K} \right],
\label{eq:int_sn_dn}
\end{equation}
where we used $\ov{\cal E}(4\ov{\sf K}) = 4\,\ov{\sf E}$.

\subsection{Jacobi Zeta and Theta Functions}

We now introduce the Jacobi zeta function \cite{Lawden_1989,NIST_Jacobi} 
\begin{eqnarray}
{\cal Z}(z,m) &\equiv& {\cal Z}\left(\phi(z|m)\;|\; m\right) \nonumber \\
 &=& {\cal E}(z|m) \;-\; \frac{{\sf E}(m)}{{\sf K}(m)}\;z,
 \label{eq:Z_def}
\end{eqnarray}
where the Jacobi amplitude function $\phi(z|m)$ is defined by the relation $\sin\phi(z|m) = {\rm sn}(z|m)$. If we substitute the definition \eqref{eq:Z_def} into Eq.~\eqref{eq:int_sn_cn_zeta}, we obtain
\begin{eqnarray}
3\,m^{2}\int_{0}^{\zeta}{\rm sn}^{2}u\;{\rm cn}^{2}u\,du &=&  \frac{3\,J\,\theta}{4\,\nu^{3}} + (2 - m)\,{\cal Z}(\zeta,m) \nonumber \\
 &&+\; \frac{1}{2}\,\pd{}{\zeta}{\rm dn}^{2}(\zeta|m),
\label{eq:Jacobi_zeta}
\end{eqnarray}
where we used the definition \eqref{eq:J_I} and 
\[ -\;m\;{\rm sn}\,\zeta\;{\rm cn}\,\zeta\;{\rm dn}\,\zeta \;=\; \frac{1}{2} \pd{}{\zeta}{\rm dn}^{2}\zeta. \]
If we substitute the definition \eqref{eq:Z_def}  into Eq.~\eqref{eq:int_sn_dn_xi}, on the other hand, we obtain
\begin{eqnarray}
\int_{0}^{\xi}\ov{\rm sn}^{2}u\;\ov{\rm dn}^{2}u\,du &=& \frac{\ov{J}\,\ov{\theta}}{4\,\nu^{3}\sqrt{m}} + \frac{1}{3}(2 - m)\,{\cal Z}(\xi,\ov m) \nonumber \\
 &&+\; \frac{1}{6}\,\pd{}{\xi}\ov{\rm cn}^{2}(\xi|\ov{m}),
 \label{eq:Jacobi_xi}
\end{eqnarray}
where we used the definition \eqref{eq:J_II} and we used
\[ -\;\ov{\rm sn}\,\xi\;\ov{\rm cn}\,\xi\;\ov{\rm dn}\,\xi \;=\; \frac{1}{2} \pd{}{\xi}\ov{\rm cn}^{2}\xi. \]

We note that Eqs.~\eqref{eq:Jacobi_zeta} and \eqref{eq:Jacobi_xi} may be expressed as partial derivatives of periodic theta functions \cite{Lawden_1989} if we express the Jacobi zeta function as
\begin{equation}
{\cal Z}(z,m) \;=\; \pd{}{z}\ln\vartheta_{4}\left(\frac{\pi z}{2{\sf K}(m)}, {\sf q}(m)\right),
\label{eq:zeta_theta}
\end{equation}
and the Jacobi elliptic functions
\begin{eqnarray}
{\rm dn}^{2}(z|m) &=& \sqrt{1 - m}\;\frac{\vartheta_{3}^{2}(\pi z/2{\sf K}, {\sf q})}{\vartheta_{4}^{2}(\pi z/2{\sf K}, {\sf q})}, \label{eq:dn_theta} \\
{\rm cn}^{2}(z|\ov m) &=& \sqrt{m - 1}\;\frac{\vartheta_{2}^{2}(\pi z/2\ov{\sf K}, \ov{\sf q})}{\vartheta_{4}^{2}(\pi z/2\ov{\sf K}, \ov{\sf q})},
\label{eq:cn_theta}
\end{eqnarray}
where $\vartheta_{2}(u,\ov{\sf q}) = 2\sum_{n = 0}^{\infty} \ov{\sf q}^{(n + \frac{1}{2})^{2}}\,\cos[(2n + 1) u]$ denotes another Jacobi theta function.

\end{document}